\documentclass{ws-procs9x6}

\newcommand{\be}{\begin{eqnarray}}
\newcommand{\ee}{\end{eqnarray}}
\begin{document}

\title{ Novel hard  semiexclusive processes and 
color singlet clusters in hadrons.}
\author{Leonid Frankfurt}
\address{Physics Department, Tel Aviv University, Tel Aviv, Israel}
\author{M.V. Polyakov}
\address{Institut f\"ur Theoretische Physik II, Ruhr--Universit\"at Bochum,\\
 D--44780 Bochum, Germany\\}

\author{M. Strikman}
\address{Department of Physics, Pennsylvania
State University, University Park, PA 16802, USA}

\author{D.Zhalov}
\address{Department of Physics, Pennsylvania
State University, University Park, PA 16802, USA}

\author{M.Zhalov}
\address{Petersburg Nuclear Physics Institute, Gatchina, Russia}

\maketitle
\abstracts{
Hard scattering to a three cluster
 final state is suggested as a method
to probe configurations in hadrons containing small size 
color singlet cluster and a residual quark-gluon system of 
a finite mass.
Examples of such processes include 
$e + N \rightarrow e+ p +M_X (\Lambda+M_X'), p+p \rightarrow
p+p+M_X(p+\Lambda+M_X')$ where $M_X(M_X')$ 
could be a pion(kaon) or other state of finite 
mass which does not increase with momentum transfer ($Q^2$). 
We argue that different models of the nucleon
 may lead
 to very different qualitative predictions for the spectrum 
of states $M_X$.
We find  that in the pion model of nonperturbative 
$q \bar q$ 
sea in a nucleon the 
cross section of these reactions is comparable to 
the cross section of the corresponding two-body reaction.
Studies of these reactions are feasible using both fixed target 
detectors (EVA at BNL, HERMES at DESY)  and collider detectors
with a good acceptance in the forward direction.}

\section{Introduction}
The QCD analyzes of the eighties have demonstrated
 that cross sections 
of hard two-body exclusive processes are  expressed through 
the minimal Fock space
components of hadrons involved in the reaction (for recent reviews and references see \cite{Rad,Brodsky2000}). 
The need to establish at what 
$Q^2$ minimal Fock space 
components start to dominate 
in these processes
has stimulated searches for the color transparency phenomena.

It is natural to move one step further and 
ask a question whether collapsing of three valence quarks 
to a small size color singlet configuration in a nucleon or 
of valence quark and antiquark
 in a meson would result in 
disappearance of other constituents?
Such scenario would be natural in quantum electrodynamics 
for the case of positronium - the 
photon field disappears in the case when 
electron and positron are close together
(that is the ratio of the amplitude for positronium 
to be in the Fock state $e^+e^-\gamma$ and 
in the Fock state $e^+e^-$ decreases with decrease of 
of the distance  between $e^+$ and $e^-$).
However  in QCD where interactions at large 
distances are strong
it is possible that non minimal Fock components of the 
hadron contain configurations with small 
color singlet  clusters without an additional smallness as
compared to  the minimal Fock component.
In fact there is no need to restrict the question to the
case of clusters build of the valence quarks in a hadron - 
one can as
well consider any small color singlet
cluster - for example a $q\bar q$
 cluster in the nucleon, a three quark cluster with 
strangeness or charm in the nucleon, etc.

In this talk we will consider the knock out of the clusters induced both by
electrons and by hadrons\footnote{As far as we know these processes were first discussed in \cite{baryon} where they were referred to as star dust processes.}.

\section{Exclusive production of forward baryons off nucleons}
\label{sec2}

Over the last few years $q\bar q$ clusters in the
nucleons and mesons were implicitly considered in
the context of the study of exclusive DIS processes:
$\gamma^*_L+N \to ``meson'' + baryon''$  for
which the factorization theorem is valid \cite{BFGMS,CFS}
which states that 
 in the 
limit of large $Q^2$ the amplitude of the 
process  at fixed $x$ 
is factorized into the convolution of a hard interaction block
calculable in perturbative QCD,  the  short-distance 
$q\bar q $ wave function of the meson, and the generalized/skewed
parton distribution (GPD) in the nucleon. The HERA data have confirmed
a number of the key predictions of \cite{BFGMS} 
including shrinkage of the $t$-distribution the light meson
production with increase of $Q^2$ to the limiting value which is close to the
slope of the $J/\psi$ production cross section. The $Q^2$ dependence of
the slope is consistent with predictions of \cite{FKS96} (which extended the
analysis of \cite{BFGMS} to account for the geometrical higher twist effects
which arise due to the finite transverse size of the virtual photon wave function)
indicating that at $Q^2\ge 4 GeV^2$ small transverse size configurations in $\rho$-meson are selected and that the suppression of the color dipole - nucleon
interaction occurs already for transverse distances $\le 0.4 $ fm.

The proof of the  factorization for the meson exclusive production \cite{CFS},
is essentially based on the observation that the
  cancellation of the soft gluon interactions
is
intimately related to the fact that
the meson arises from a
quark-antiquark pair generated by the hard scattering.  Thus the
pair starts as a small-size configuration and only substantially
later grows to a normal hadronic size, to a meson. 
Similarly,
the  factorization theorem should be  valid
 for  the  production of leading baryons
\begin{equation}
     \gamma ^{*}(q) + p \to  B(q+\Delta ) + M(p-\Delta ),
\label{process1}
\end{equation}
and even leading antibaryons
\begin{equation}
     \gamma ^{*}(q) + p \to  \bar{B}(q+\Delta ) + B_2(p-\Delta ),
\label{process2}
\end{equation}
where $B_2$ is a system with the baryon charge of two.
For example in the case of the process \ref{process1} the dominant diagram is
given by Fig.1:
\begin{figure}
    \begin{center}
        \leavevmode
        \epsfxsize=.5\hsize
        \epsfbox{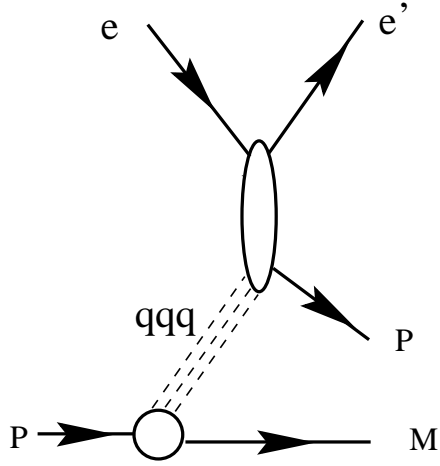}
    \end{center}
\caption{\it Production of a fast baryon  and recoiling mesonic system.}
\end{figure}

In QCD to describe the hard exclusive processes one needs to use 
generalized (skewed) parton distributions. 
Since the objects one has to introduce for 
description of  $N\to N$ transitions and non-diagonal
transitions like $N\to \Lambda, \Delta$ are pretty different
we suggest to
refer to the first type of distributions as 
 as  generalized parton distributions (GPD), while  in the case of 
non-diagonal transitions use the term skewed PD.

To describe in QCD  process
(\ref{process1}),  one needs to introduce a new
 non-perturbative mathematical
object \cite{FPS}
 in addition to the GPDs and SPDs.
It can be called a super skewed
parton distribution (amplitude).
It is
 defined as
a non-diagonal matrix element of the tri-local quark operator
between a meson $M$ and a proton:
\be
\nonumber
&&\int \prod_{i=1}^3 dz_i^- \exp[i\sum_{i=1}^3 x_i\, (p\cdot z_i)]\cdot \,
\nonumber \\
&&\langle M(p-\Delta)|\varepsilon_{abc}
\, \psi^{a}_{j_1}(z_1) \,
\psi^{b}_{j_2}(z_2) \,
\psi^{c}_{j_3}(z_3)|N(p)\rangle\Bigl|_{z_i^+=z_i^\perp=0}
\nonumber \\
&&=\delta(1-\zeta-x_1-x_2-x_3)\, F_{j_1 j_2 j_3}(x_1,x_2,x_3,\zeta,t) \;,
\ee
where $a,b,c$ are color indices, $j_i$ are spin-flavor indices, and 
$F_{j_1 j_2 j_3}(x_1,x_2,x_3,\zeta,t)$ are the  new  superSPDs. They 
can be decomposed into
invariant spin-flavor structures
which depend on the quantum numbers of the meson $M$.
They
depend on the variables $x_i$ (which are contracted with
the  hard kernel in
the amplitude), on the skewedness parameter $\zeta=1-\Delta^+/p^+$
and the momentum transfer squared $t=\Delta^2$.
In some sense, with this definition of $\zeta$ the limit
$\zeta\to 0$ corresponds to the usual distribution amplitude, $i.e.$
skewedness $\to 0$ means (for appropriate quantum numbers of the current)
superSDP $\to$ ``nucleon distribution amplitude''.

Though quantitative
calculations of  processes (\ref{process1}, \ref{process2})
 will take  time, some
qualitative predictions could be made
right away. First we observe that in the Bjorken limit
the light cone fraction of the slow meson satisfies condition:
\begin{equation}
\alpha_h={p_{M-}\over p_{N-}}={E_h-p_{3M}\over E_N-p_{N3}}=
{E_M-p_{3M}\over m_N}=1-x
\end{equation}
and its transverse momentum $p_t$ relative to the $\vec q$ direction been
fixed. To ensure an early onset of scaling it is natural to consider
 the process as a function of $Q^2$ for fixed $\alpha_h, p_t$. This way 
we can make a natural link to the picture of removing  a cluster
 from the nucleon leaving the residual system undisturbed.
If the color transparency suppresses the final state interaction
between the fast moving nucleon and the residual meson state
 early enough it would be natural to expect an early onset of 
the factorization
of the cross section to a function 
which depends on $\alpha_h,p_t$ and the cross section 
of the electron-nucleon elastic scattering:
\begin{equation}
{d\sigma(e+N\to e+N + M) \over
{d\alpha_Md^2p_t/\alpha_M}}=f_M(\alpha_M,p_t) (1-\alpha_M) \sigma(eN\to eN),
\label{escale}
\end{equation}
where $(1-\alpha_M)$ is the flux factor and $ \sigma(eN\to eN)$ in 
the cross section of the elastic $eN$ scattering in the appropriate kinematics.

In the case of the pion production the soft pion limit corresponding to 
$1-x \sim m_{\pi}/m_N, p_t \le m_{\pi}$
 is of special interest because one could use
the factorization theorem and the  chiral perturbation 
theory similar to consideration of the process
$eN\to eN\pi$ at large $Q^2$ and small W \cite{PPS}. However
reaching this kinematics would require extremely high $Q^2$.
At the same time reactions with leading nucleon
for $x\le 0.3 $ could be studied at sufficiently large $W$ already at 
Jlab and HERMES.

Reaction (\ref{process1}) provides also 
a promising avenue
to look for exotic meson states including gluonium.
Indeed, if one would consider, for example, the MIT bag model, 
the  removal of
 three quarks from the system  could leave the
residual system looking    like a bag made predominantly  of glue.
It is natural to expect that such a system would have a large overlapping
integral with gluonium states.

An interesting example of a related process where cluster structure can 
manifest itself is the 
deep inelastic exclusive diffraction at large enough t:
\begin{equation}
e+p \rightarrow e+ leading ~\rho + M +B
\label{HERA}
\end{equation} 
where $ -t=-(p_{\gamma^*}-p_{\rho})^2 \ge 2 GeV^2$, $Q^2\ge few ~ GeV^2$
and $p_t(M) \approx p_t(\rho) $ and $ p_t(M) \gg  p_t(B)$. 
The $t$- dependence of the process $e+p \rightarrow e+  ~\rho + p$ is
 predominantly determined by the two-gluon 
nucleon form factor and it can be fitted as $\approx 1/(1-t/m^2)^4
$ with $m^2 \sim 1 GeV^2$ at intermediate energies and 
 $m^2 \sim 0.6 GeV^2$ at HERA energies. In the case of scattering off a
 meson a natural guess would be that the t-dependence is much slower -
perhaps  $\approx 1/(1-t/m^2)^2$ with similar $m^2$.
So for large enough $t$ this process may have cross section comparable to the 
cross section of the $e+p \rightarrow e+  ~\rho + p$
process. Study of this process could provide a test of the interpretation
of the smaller gluon radius of the nucleon indicated by our recent
 analysis \cite{FS2002}.

\section{Hadron induced hard semiexclusive processes}

A natural extension of the processes discussed for electron scattering is
hadron scattering process:
\begin{equation}
A+B\to C_{int}+C_{sp} + D,
\label{2to3}
\end{equation}
where sufficiently large momentum is transfered to $C_{int}$ and $D$
(scattering at finite c.m. scattering angles in the c.m. frame of $C_{int}$
 and 
$D$), while $C_{sp}$ similar to the case of 
the process \ref{process1} is produced in the fragmentation region of either
$A$ or $B$.

Taking for certainty $D$ in the target fragmentation of $B$
we can expect that in the color transparency approximation the process
will proceed via scattering of a hadron $A$ in the minimal Fock
 space configuration off 
 a color singlet cluster in the hadron $B$ with minimal number 
of constituents allowed for the process $A+ ``cluster'' \to 
C_{int} + D$.
An obvious practical advantage of these processes as compared to
the processes \ref{2to3} is that one can use different beams - pions, 
kaons, hyperons to probe the clustering structure of different hadrons
while the processes \ref{process1} in practice are restricted to the 
case of proton targets.

The two-body large angle hadron scattering processes are known to 
satisfy to a good approximation dimensional counting rules, for a review see
\cite{Stan}. At sufficiently large momentum transfer the small size 
configurations should give the dominant contribution and hence the
 rescattering effects should be small.
Hence we expect the scaling relations for these processes of the similar kind
{\it for a fixed value of $\alpha_{C_{sp}},p_{t~~C_{sp}}$}:
\begin{equation}
{d\sigma(A+B\to C_{int}+ C_{sp} + D) \over
{d\alpha_{sp}d^2p_{t~sp}/\alpha_{sp}}}=
\phi(\alpha_{sp},p_{t~sp}) R(\theta_{c.m.})\left({s_o/s'}\right)^n
\label{hscale}
\end{equation}
where $s'=(p_{C_{int}} +p_D)^2$,  $\theta_{c.m.}$
 is the c.m. angle in the 
$C_{int}-D$ system, and $n$
 is expressed 
through the number of constituents  involved in the subprocess
in the same way as in the two-body large angle scattering:
\begin{equation}
n=n_q(A)+n_q(cluster)+n_q(C_{int})+n_q(D) - 2.
\end{equation}

There is a number of 
the processes where the hard subprocess resembles the scattering off
two hadrons for which the cross section is known, like the process
$p+p\to p+p + M_{spect}$, or $p+p\to p+\pi + N_{spect}$ presented in Fig.2.
\begin{figure}
    \begin{center}
        \leavevmode
        \epsfxsize=.6\hsize
        \epsfbox{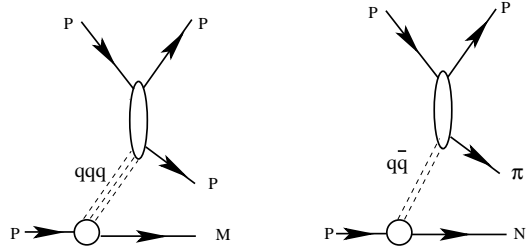}
    \end{center}
\caption{\it (a) Production of two high $p_t$ baryons
 and recoiling mesonic system,
(b) production of a high $p_t$ pion and a nucleon and recoiling baryonic system.}
\end{figure}

In this case 
we can write an interpolation formulae 
similar to the ones for the electron scattering:
For example, 
\begin{equation}
{d\sigma(p+p\to p+p + \pi^0) \over
d\alpha_{\pi^0}d^p_t/\alpha_{\pi^0}}= F(\alpha_{\pi},p_t) (1-\alpha_pi)
d\sigma^{pp\to pp}(s',\theta_{c.m.})
\end{equation}
with $s'\approx =(1-\alpha_{\pi})s$. Since the hard cross section decreases 
strongly with increase of $s'$ at fixed $\theta_{c.m.}$ one expects a strong 
enhancement of production of mesons with relatively high values of 
$\alpha_{\pi}$
in these processes. Also, assuming that the distributions of  the 
small color clusters contributing to the electron reaction and to the 
hadron reactions are about the same we can get scaling relations between
 the cross section of proton and electron induced processes.  
For example, 
\begin{equation}
{{d\sigma(p+p\to p+p + \pi^0) \over
d\alpha_{\pi^0}d^2p_t/\alpha_{\pi^0}}
\over 
{d\sigma(e+N\to e+N + \pi^0) \over
{d\alpha_{\pi^0}d^2p_t/\alpha_{\pi^0}}}} \approx 
{\sigma(p+p\to p+p)\over \sigma(eN\to eN)},
\label{pscale}
\end{equation}

We have performed first estimates of the rate of the production 
of pions using the discussed mechanism and using a simple model for
the $\Gamma_{NN\pi}$ vertex. Enhancement of the scattering off three
 quark clusters which carry $(1-x_{\pi})$ fraction of the total 
light cone momentum of the nucleon as compared to the scattering off the 
nucleon as a whole by a factor $(1-x_{\pi})^{-10}$ leads to enhanced role
 of the pion cloud and results in the cross sections of the same magnitude
 as the elastic $pp$ scattering. We are currently performing  more detailed
studies to determine the contribution of this mechanism to the cross section
 measured by EVA \cite{Heppelman}.

We also found that at the intermediate energies E$\le$ 10 GeV studied at
 EVA \cite{Heppelman}
an important  background to the discussed mechanism
 is provided by production
of excited nucleon states. Indeed, the 
cross section of the processes like $pp\to N^*+p$ at 
$\theta_{c.m.}\sim 90^0$  is comparable to the elastic $pp$ scattering.
Moreover if we would sum over  a sufficient range of masses of $N^*$'s 
we should expect to find a cross section which is 
enhanced as compared to the elastic scattering 
since it does not contain  a smallness for  three quarks  to 
transform to a particular state (nucleon). If we use the quark
counting rules as a guide, the ratio of quasielastic and elastic
processes
should increase  with $s$  as  $\propto s^2$.  At sufficiently
high energies kinematics of the meson production in these processes
is  qualitatively different from the process \ref{2to3} since 
the average  pion 
transverse momenta in this case increase with $s$ approximately
as $\propto \sqrt{s}$ (for the case of the $N\pi$ final state). 
However we have checked that for $E_{inc}\le 10 GeV$ 
it is difficult to separate  fragmentation
 in the initial and final states even in the simplest reaction 
$pp\to pp \pi^0$  except for production of pions at very small $p_t$.

In the previous discussion we assumed that  hard two body
subprocess is dominated by  the scattering of constituents
in small size configurations, so that the interaction with residual
system could be neglected. However the current data on transparency in
high-energy large angle $(p,2p)$ reactions suggests  a rather
complicated interplay of the contributions of large and small size
configurations. In this case the initial and final
state interactions with the would be spectator would be possible.
For example in the process $pp\to pp \pi^0$ with $\pi^0$ a spectator
both the proton in the initial state and both protons in the final
state can rescatter off the pion.
Presence of multiple rescatterings will lead to rather complicated
patterns of angular correlations similar to those found for the 
large angle $p^2H\to ppn$ process in \cite{pdppn}.
These rescatterings will weakly affect distribution over $\alpha_M$ 
though they would
change 
overall absolute value of the 
cross section and lead to broadening of the $p_t$ distribution of the
spectator.

Another possible source of angular asymmetries is postselection of the initial 
state which could be different at intermediate energies when the size of 
three quark cluster is not too small. The requirement of the 
elastic scattering off the three quark cluster may select the alignment 
of the cluster relative to the reaction axis hence modifying the 
angular distribution of the spectator system. 
In particular this effect can emerge
 as a kind of a hadron level Sudakov radiation, see discussion in
 \cite{Heppelman}.

To summarize,
a systematic study of the lepton and hadron
induced hard semiexclusive reactions
with leading baryons  is necessary. It would provide a 
qualitatively new information 
 about correlations of partons in hadrons and 
as well as about the 
dynamics of the large angle elastic scattering.

\section{Acknowledgments}
This work has been supported in part by the 
USDOE, the Humbold foundation, GIF, NSF, CRDF.

\end{document}